\documentclass[10pt]{iopart}

\usepackage{iopams}  
\newcommand{\beq}{\begin{equation}}
\newcommand{\eeq}{\end{equation}}
\newcommand{\beqa}{\begin{eqnarray}}
\newcommand{\eeqa}{\end{eqnarray}}
\newcommand{\kB}{\mbox{$k_{\rm B}$}}
\newcommand{\kBT}{\mbox{$k_{\rm B}T$}}
\newcommand{\TR}{\mbox{$T_{\rm R}$}}

\usepackage{graphicx}

\begin{document}

\title[Phase diagram of step faceting for sticky steps]{Phase diagram of step faceting for sticky steps}

\author{Noriko Akutsu}

\address{Faculty of Engineering, Osaka Electro-Communication
 University, Neyagawa, Osaka 572-8530, Japan}
\ead{nori3@phys.osakac.ac.jp}
\vspace{10pt}
\begin{indented}
\item[]\today
\end{indented}

\begin{abstract}

A phase diagram for the step faceting phase, the step droplet phase, and the Gruber-Mullins-Pokrovsky-Talapov (GMPT) phase on a crystal surface is obtained by calculating the surface tension with the density matrix renormalization group method. 
The model based on the calculations is the restricted solid-on-solid (RSOS) model with a point-contact-type step-step attraction (p-RSOS model) on a square lattice.
The point-contact-type step-step attraction represents the energy gain obtained by forming a bonding state with orbital overlap at the meeting point of the neighbouring steps.
Owing to the sticky character of steps, there are two phase transition temperatures, $T_{f,1}$ and $T_{f,2}$.
At  temperatures $T < T_{f,1}$, the anisotropic  surface  tension has a disconnected shape around the (111) surface. 
At $T<T_{f,2}<T_{f,1}$, the surface  tension has a disconnected shape around the (001) surface.
On the (001) facet edge in the step droplet phase, the shape exponent normal to the mean step running direction $\theta_n=2$ at $T$ near $T_{f,2}$, which is different from the GMPT universal value $\theta_n=3/2$.  
On the (111) facet edge, $\theta_n=4/3$ only on $T_{f,1}$.  
To understand how the system undergoes phase transition, we focus on the connection between the p-RSOS model and the one-dimensional  spinless quasi-impenetrable attractive bosons at absolute zero.

\end{abstract}

%
%
\submitto{\JPCM}
%
%
%


\section{Introduction}

In 1961, Mullins showed that faceted steps are formed during crystal growth \cite{mullins}.
The faceting of steps has attracted much attention in the study of the morphology of crystal surfaces. 
Controlling step faceting is important for obtaining high-quality crystals in industry, and controlling crystal morphology is also important for nanotechnology.

\begin{figure}
\centering
\includegraphics[width=8 cm,clip]{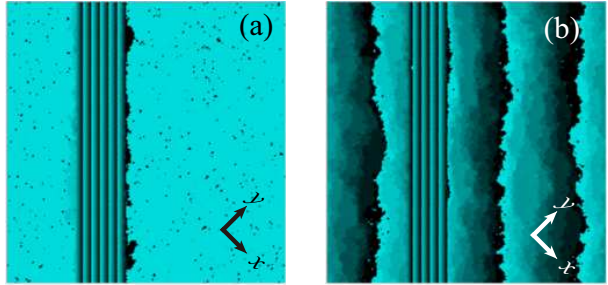}%
\caption{
Top view of vicinal surfaces calculated by a Monte Carlo method.
The surfaces are inclined in the $\langle 111 \rangle $ direction.
The surface height is represented by brightness with 10
gradations, with brighter regions being higher.
$\epsilon_{\rm int}/\epsilon=-1.3.$
Time: 2$\times 10^8$ Monte Carlo steps per site (MCS).
Size: $160\sqrt{2} \times 160\sqrt{2}$.
(a) $\kBT/\epsilon=0.8$.  Number of steps= 60.
 (b) $\kBT/\epsilon=0.83$. Number of steps= 80.
}
\label{MC}
\end{figure}

The step faceting occurs not only on a growing surface but also on a crystal surface near equilibrium (figure  \ref{MC}). 
Cabrera \cite{cabrera} studied the relationship between the surface free energy of the crystal and the stability of a vicinal surface phenomenologically, where the vicinal surface was tilted from a low Miller-index surface. Type I and type II anisotropy was assumed in the surface free energy, and the instability of the macro-step formation with type II anisotropy in the surface free energy on the vicinal surface was discussed.
Cabrera also investigated the equilibrium crystal shape (ECS), which is the shape of the crystal droplet with the minimum surface free energy. 
For type II anisotropy in the surface free energy, the ECS has a facet with a sharp facet edge \footnote{Recently, the sharp facet edge on the ECS has been called the first-order shape transition.}, where the surface slope jumps from zero to a finite value. 
However, calculations based on a microscopic model that causes type II anisotropy in surface free energy were not reported.


\begin{figure}
\centering
\includegraphics[width=8 cm,clip]{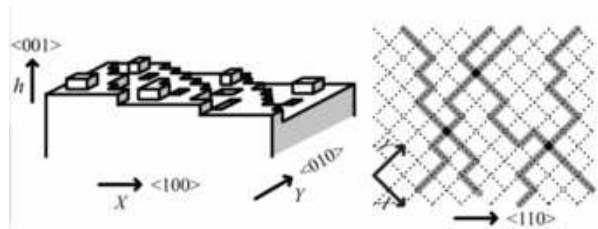}%
\caption{
(a) Perspective view of a vicinal surface inclined in the $\langle 111\rangle$ direction in an RSOS model.
$X=\eta_x/\kBT=-\lambda x/\kBT$, where $\lambda$ is the Lagrange multiplier relating to the crystal volume \cite{andreev}.
$Y=\eta_y/\kBT=-\lambda y/\kBT$.
(b) Top view of a vicinal surface inclined in the $\langle 111\rangle$ direction.
Thick gray lines represent surface steps.
Closed squares represent meeting points between adjacent steps. 
}
\label{vicinal}
\end{figure}


In our previous work  \cite{akutsu10}-\cite{akutsu14}, we presented a restricted solid-on-solid (RSOS) model on a squared lattice (figure  \ref{vicinal}) with a point-contact-type step-step attraction RSOS (p-RSOS) model.
Here, ``restricted'' means that the height difference between nearest neighbour (nn) sites is restricted to $\{0, \pm 1\}$.
We consider that the origin of the point-contact-type step-step attraction is the orbital overlap of the dangling bonds at the meeting point of the neighbouring steps.
The energy gained by forming the bonding state is regarded as the attractive energy between steps.

From numerical calculations using the density matrix renormalization group (DMRG) method \cite{dmrg}-\cite{schollwock}, we obtained the azimuthal dependence of the surface tension, which is disconnected \cite{akutsu12}.
As a result of the disconnected surface tension, a faceted merged step is formed on a vicinal surface at low temperatures as the two surfaces coexist. 
We showed that the macro-step faceting occurs on the vicinal surface accompanied by the first-order shape transition on an ECS \cite{akutsu10}-\cite{akutsu14}.


\begin{figure}
\centering
\includegraphics[width=8.5 cm,clip]{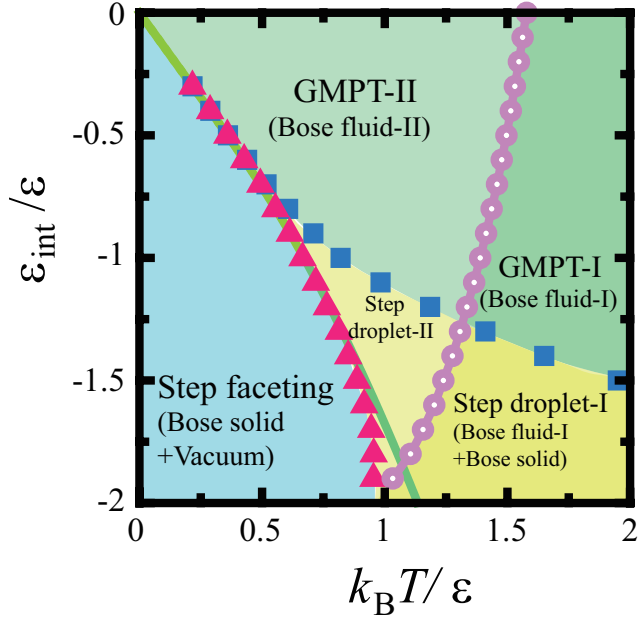}%
\caption{
Phase diagram of p-RSOS model for a vicinal surface obtained by the PWFRG method.
Triangles: calculated first-order phase transition points ($T_{f,2}$).
Squares: calculated points $T_{f,1}$.
Open circles: calculated roughening transition temperatures of the (001) surface.
Solid line: equation (\ref{kBTf2})
}  
\label{phase_diagram}
\end{figure}

In this paper, we present the following work.
1) We construct a phase diagram (figure  \ref{phase_diagram}) for the step faceting phase (one-dimensional (1D) Bose solid + vacuum), the step droplet (1D Bose liquid + solid) phase, and the Gruber-Mullins-Pokrovsky-Talapov (GMPT \cite{gmpt}; 1D Bose fluid)  phase  obtained by numerical calculations.
2) We give the expanded form of the surface free energy density $f(\vec{p})$ with respect to $|\vec{p}|$ as
\beqa
f(\vec{p})&=&f(0)+\gamma \frac{|\vec{p}|}{d_1}+ A \frac{|\vec{p}|^2}{d_1^2}+ B \frac{|\vec{p}|^3}{d_1^3} 
+ C_4  \frac{|\vec{p}|^4}{d_1^4} 
\nonumber \\
&&
+ {\mathcal O}(|\vec{p}|^5),\label{f_final}
\eeqa
where $\vec{p}$ is the surface gradient, such as $\vec{p}= (\partial z(x,y)/\partial x, \partial z(x,y)/\partial y)$, $z(x,y) $ is the profile of the surface, $d_1=1$ is the height of a single step, $\gamma$ is the step tension, $A$ is the step-coalescence factor, $B$ is the step-interaction coefficient, and $C_4$ is the step-collision coefficient.
To understand how the phase transition occurs and how the coefficients are determined, we focus on the connection between the surface model and the 1D quantum particle systems at absolute zero.

Historically, in the study of the roughening transition and the faceting transition \cite{beijeren}-\cite{einstein2015book}, the surface free energy density of a vicinal surface with step-step repulsion has the GMPT universal form \cite{gmpt}.
At temperatures less than the roughening transition temperature of a terrace, the surface free energy density is described as
\beq
f(\rho)= f(0) + \gamma \rho + B \rho^3 + {\mathcal O}(\rho^4), \label{f-gmpt}
\eeq
where $\rho$ is the step density and $\rho=|\vec{p}|/d_1$.
The lack of the quadratic term ($A=0$) with respect to $\rho$ is known as the GMPT universal form.

The GMPT form of the surface free energy (equation  (\ref{f-gmpt})) is first confirmed by the exact calculations on the body-centred cubic solid-on-solid model \cite{beijeren,jayaprakash}.
The terrace-step-kink (TSK) models in the lattice system and the continuous system are equivalent to the  spinless free fermion (FF) system in one dimension \cite{izuyama}-\cite{yamamoto97} .
This is because the steps in the TSK model can be regarded as the {\it impenetrable} elementary excitations in two dimensions.

Furthermore, the TSK model, with a long-range step-step repulsion of $g/(x_i-x_i+1)^2$, where $g$ is the coupling constant and $x_i$ is the location of the i-th step,  has a GMPT form (equation  (\ref{f-gmpt})) \cite{jayaprakash84-2,williams93,yamamoto94} based on the exact results of Calogero \cite{calogero} and Sutherland \cite{sutherland}.
This long-range step-step repulsion is generated by the elastic interaction between steps \cite{marchenko,alerhand}.

For the Si(111) surface, Williams {\it et al.} \cite{williams93} showed that step bunching occurs because of competition between the polymorphic surface free energy with respect to the (1$\times$1) surface structure and the  (7$\times$7) reconstructed surface structure. 
The respective surface free energies have the GMPT form equation  (\ref{f-gmpt}).
They also estimated $g$ of Si(111) by measuring surface quantities.  

In the p-RSOS model, the vicinal surface inclined in the $\langle 111 \rangle$ direction caused the phase transitions shown in figure  \ref{phase_diagram}.
As a result of the special geometry of RSOS models on a squared lattice, two neighbouring steps can occupy one site at a time, but more than two steps cannot occupy one site at a time.
We call this special character {\it quasi-impenetrability}.
Due to the quasi-impenetrability, the steps on the RSOS model inclined in the $\langle 111 \rangle$ direction should be mapped to bosons in one dimension. 

However, the vicinal surface inclined in the $\langle 101 \rangle$ direction does not cause such phase transitions.
This suggests the Hamiltonian of the coarse-grained vicinal surface inclined in the $\langle 111 \rangle$ direction is different from the Hamiltonian of the coarse-grained vicinal surface inclined in the $\langle 101 \rangle$ direction.
This difference should affect the $|\vec{p}|$-expanded form of the surface free energy density in equation  (\ref{f_final}).

The 1D interacting boson system also has attracted attention because recent experiments with ultra-cold quantum gases \cite{dettmer,kinoshita,parades} have created a quasi-1D interacting boson system.
Controlled repulsive or attractive interactions among atoms in the 1D array in the experiments have triggered new theoretical studies of 1D interacting bosons \cite{cazalilla}, indicating that 1D interacting boson system are relevant in a wide variety of contexts for condensed matter physics.

This paper is organized as follows.
In \S \ref{sec_model0}, we explain the p-RSOS model and calculation method.
Examples of the calculated Andreev free energy and surface tension are shown.
In \S \ref{sec_phase_diagram}, the phase diagram for the crystal surface is explained.
The effect of the sticky character of steps on surface roughness is also discussed.
  The difference between the vicinal surface inclined in the $\langle 111 \rangle$ direction and the vicinal surface inclined in the $\langle 101 \rangle$ direction is discussed in \S \ref{sec_1DQP}.
In \S \ref{sec_1Dattractive_bosons}, the phase sepalation line (PSL) between the step faceting phase and step droplet phase (lower PSL) is explained as the solidifying of bosons.
The PSL between the step droplet phase and the GMPT phase (upper PSL) is discussed with respect to the melting of Bose solid.
Non-GMPT shape exponents are discussed in this section.
The discussion and conclusions are presented in \S \ref{sec_discussion} and \S \ref{sec_conclusions}, respectively.

\section{\label{sec_model0}p-RSOS model}

\subsection{\label{sec_model} The model}

In this section, we introduce the surface model, and then explain how the surface free energy and other thermodynamic quantities are obtained numerically.

The surface undulation is described by an RSOS model \cite{rsos} with a surface height of $h(\vec{r})$ at site $\vec{r}=(x,y)$ on the square lattice.
The height differences between nn sites costs the energy $\epsilon |h(\vec{r})-h(\vec{r'})|$, where $\epsilon$ is the microscopic ledge energy.
Here, ``restricted'' means that the height differences between nn sites are restricted to $\{0,\pm 1\}$.

The Hamiltonian of the p-RSOS model \cite{akutsu10}-\cite{akutsu14} is  written as
\beqa
{\mathcal H}_{\rm p-RSOS} &=& \sum_{i,j} \epsilon 
[ |h(i+1,j)-h(i,j)|   \nonumber \\
&&
+|h(i,j+1)-h(i,j)|]   \nonumber \\
&& +\sum_{i,j} \epsilon_{\rm int}[ \delta(|h(i+1,j+1)-h(i,j)|,2)  \nonumber \\
&&
 +\delta(|h(i+1,j-1)-h(i,j)|,2)] ,   \label{hamil}
\eeqa
where $\delta(a,b)$ is Kronecker's delta.
The summation with respect to $(i,j)$ is performed all over sites on the square lattice.
The RSOS restriction is required implicitly.
For $\epsilon_{\rm int}<0$, the interaction among steps becomes attractive.

For the vicinal surface, to make surface tilted we add the external field terms 
\beqa
{\mathcal H}_{\rm vicinal}= {\cal H}_{\rm p-RSOS}&-&\sum_{i,j}\left \{ \eta_x [h(i+1,j)-h(i,j)] \right. \nonumber \\
&+& \left. \eta_y [h(i,j+1)-h(i,j)] \right \}, \label{hamil_vicinal}
\eeqa
where $(\eta_x,\eta_y)$ is the Andreev field, which tilts the surface.

\subsection{\label{sec_z} Calculation of the partition function}

The partition function, ${\cal Z}$, for the p-RSOS model is given by
\beq
{\cal Z}= \sum_{\{h(i,j)\}} e^{-\beta {\cal H}_{\rm vicinal}} \label{partition}
\eeq
where $\beta= 1/\kBT$, $\kB$ is the Boltzmann constant, and $T$ is the temperature. 
The Andreev free energy (density), $\tilde{f}(\vec{\eta})$, is the thermodynamic potential calculated from the partition function, ${\cal Z}$, by using
\beq
\beta \tilde{f}(\vec{\eta})= - \lim_{{\cal N} \rightarrow \infty} \frac{1}{{\cal N}} \  \ln {\cal Z}, \label{tildef}\\
\eeq
where ${\cal N}$ is the number of lattice points on the square lattice.

\begin{figure}
\includegraphics[width=7.5 cm,clip]{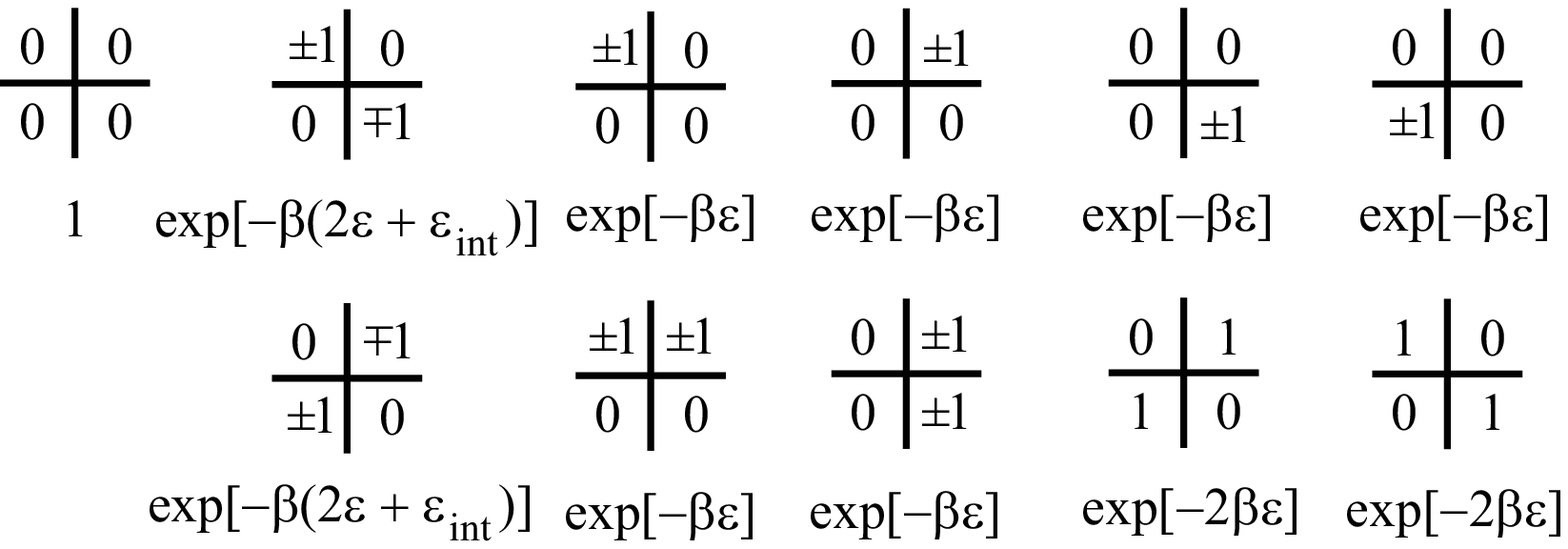}%
\centering
\caption{
Statistical weights of vertices for the p-RSOS model.
}
\label{vertex}
\end{figure}

To calculate the partition function, we adopt a transfer matrix method \cite{lieb72}.
The transfer matrix, $\hat{T}$, is obtained from the product of the vertices in the 19-vertex model \cite{akutsuJPCM11}, to which the p-RSOS model is mapped (figure \ref{vertex}).
The partition function, ${\cal Z}$ (equation  (\ref{partition})), is then rewritten in terms of  $\hat{T}$ as
\beq
{\cal Z}= {\rm Tr} [\hat{T}(t_1,t_2,\cdots,t_N;s'_1,s'_2,\cdots,s'_N)^M] \label{partition2}
\eeq
where $N$ is the number of linked vertices and $M$ is the length of the system in the vertical direction.
In the limit $M,N \rightarrow \infty$, only the largest eigenvalue of transfer matrix $\Lambda(N)$ contributes to the partition function.
Therefore, the Andreev free energy is obtained from equation  (\ref{tildef}) as
\beq
\beta \tilde{f}(\vec{\eta})= - \lim_{M,N \rightarrow \infty} \frac{1}{NM} \  \ln \Lambda(N) ^{M}. \label{tildeftm}
\eeq
The transfer matrix is diagonalized efficiently by using the product-wave-function renormalization group (PWFRG) algorithm \cite{pwfrg}-\cite{schollwock}.
The PWFRG algorithm is a variant of the DMRG method \cite{dmrg,schollwock} 
In the PWFRG calculation, the number of ``retained bases'', $m$, is set from 7 to 37.
The number of iterations for the diagonalization process is set to $200 \sim 10^4$.

\begin{figure}
\centering
\includegraphics[width=8.5 cm,clip]{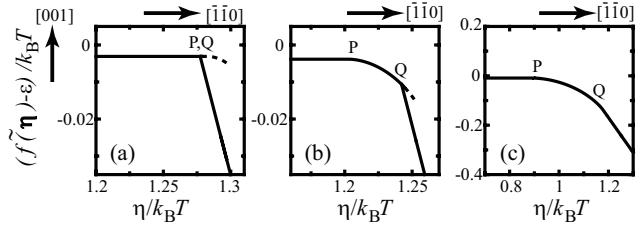}%
\caption{$\eta/\kBT$ dependence of the Andreev free energy over $\kBT$.
$\eta=|\vec{\eta}|$.
$\epsilon_{\rm int}/\epsilon = -0.9$.
Solid line: PWFRG calculations.
Broken line: metastable states.
The endpoints of the broken line represent the spinodal points approximately.
(a) $\kBT/\epsilon=0.61$.
(b) $\kBT/\epsilon=0.63$.
(c) $\kBT/\epsilon=0.72$.
$\kBT_{f,2}/\epsilon=0.6130 \pm 0.0012$.   $\kBT_{f,1}/\epsilon=0.709\pm 0.004$.
Note that  $\tilde{f}(\vec{\eta})$ is similar to the equilibrium shape of a crystal droplet (\ref{andreev-landau}).
Points P and Q show the facet edge of the (001) surface with $\eta_c^{(001)}$ and the (111) surface with $\eta_c^{(111)}$, respectively.
}
\label{ecs}
\end{figure}

Examples of the calculated Andreev free energy for $\epsilon_{\rm int}/\epsilon= -0.9$ at several temperatures are shown in figure  \ref{ecs}. 
The broken lines represent the lines for metastable states.
The end points of the respective broken lines give approximate values of the spinodal points.
Beyond the spinodal points, we cannot calculate the free energy because the states beyond the spinodal point are thermodynamically unstable.

It should be noted that the profile of the Andreev free energy, $\tilde{f}(\vec{\eta})$, as a function of the Andreev field, $\vec{\eta}$, is similar to the ECS, $z=z(x,y)$ \cite{andreev,akutsu2015book} (\ref{andreev-landau}). 
figure  \ref{ecs} shows the cross-section of the ECS, where points P and Q represent the facet edges of the (001) and (111) facets, respectively.  

figure  \ref{ecs} shows that the surface slope jumps at the (111) facet edge (point Q) from a slope of $p^*_1(T)$  to $\sqrt{2}$ for $T<T_{f,1}$, and the surface slope also jumps at the (001) facet edge (point P) from 0 to $\sqrt{2}$ for $T<T_{f,2}$.

\subsection{\label{sec_calculation}Calculation of thermodynamic quantities}

\begin{figure}
\centering
\includegraphics[width=4.5 cm,clip]{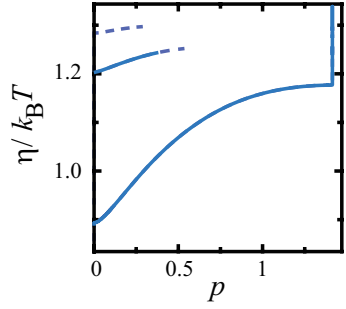}%
\caption{
$p$ {\it vs.}  $\eta/\kBT$ inclined in the $\langle 111 \rangle$  direction ($\phi= \pm \pi/4$), where $p=|\vec{p}|$ and $\eta=|\vec{\eta}|$.
$\epsilon_{\rm int}/\epsilon = -0.9$.
Solid line: PWFRG calculations.
Broken line: metastable states.
The endpoints of the broken line represent spinodal points approximately.
From top to bottom, $\kBT/\epsilon=0.61$ ($T<T_{f,2}$),  $\kBT/\epsilon=0.63$ ($T_{f,2}<T<T_{f,1}$), and $\kBT/\epsilon=0.72$ ($T>T_{f,1}$).
}
\label{eta-p-0.9}
\end{figure}

Using the PWFRG method, the surface gradient, $\vec{p}=(p_x,p_y)=\vec{p}(\vec{\eta})$, is calculated by $p_x=\langle \Delta h_x \rangle/a$ and $p_y=\langle \Delta h_y \rangle/a$ at a fixed $\vec{\eta}$, where $\Delta h_x= h(i+1,j)-h(i,j)$, $\Delta h_y= h(i, j+1) - h(i, j)$, $a$ (=1) represents the lattice constant of the square lattice, and $\langle \cdot \rangle$ represents the thermal average.
Examples of the calculated surface slopes, $p(\vec{\eta})= \sqrt{p_x^2+p_y^2}$, are shown in figure  \ref{eta-p-0.9}.

The directions $\langle 100 \rangle$ ($\phi=0$) and $\langle 110 \rangle$ ($\phi= \pi/4$) are special.
If we express $\vec{p}$ and $\vec{\eta}$ as polar coordinates, such as $(p_x,p_y)=(p \cos \phi,p \sin \phi)$ and $(\eta_x,\eta_y)=(\eta \cos v, \eta \sin v)$, respectively, $\phi \neq v$ generally.
For the directions $\langle 100 \rangle$  and $\langle 110 \rangle$, we can choose the axis so that $\phi=v$ (geodesic lines).

\begin{figure}
\centering
\includegraphics[width=8.5 cm,clip]{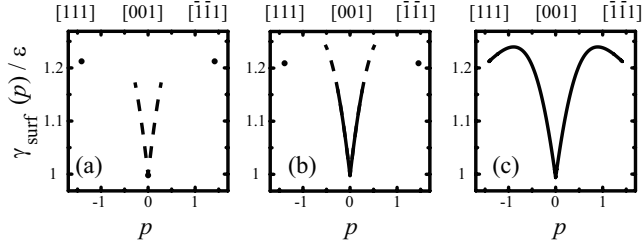}%
\caption{
$p$ dependence of surface tension inclined in the $(111)$ or $(\bar{1} \bar{1}1)$ direction ($\phi= \pm \pi/4$), where $\beta= 1/\kBT$.
$\epsilon_{\rm int}/\epsilon = -0.9$.
Solid line: PWFRG calculations.
Broken line: metastable states.
The endpoints of the broken line represent spinodal points approximately.
Closed squares: surface tension of the (111) or (001) surfaces.
(a) $\kBT/\epsilon=0.61$ ($T<T_{f,2}$).
(b) $\kBT/\epsilon=0.63$ ($T_{f,2}<T<T_{f,1}$).
(c) $\kBT/\epsilon=0.72$ ($T>T_{f,1}$).
}
\label{bgam}
\end{figure}

By using the values of the Andreev free energy (equation  (\ref{tildeftm})) and the surface gradient, $\vec{p}(\vec{\eta})$, from the PWFRG calculations, we obtain the surface free energy density, $f(\vec{p})$, from $f(\vec{p})=\tilde{f}(\vec{\eta})+ \vec{\eta}\cdot \vec{p}$ (equation  (\ref{eq_tildef_def})).
Simultaneously, the external parameter is changed from $\vec{\eta}$ to $\vec{p}$.
The surface tension, $\gamma_{\rm surf}(\vec{p})$, is obtained from the equation \cite{akutsu87}
\beq
\gamma_{\rm surf}(\vec{p})=f(\vec{p})/\sqrt{1+ p_x^2+p_y^2}.\label{eq_surface_tension}
\eeq 
In figure  \ref{bgam}, we show examples of the obtained surface tension.

\section{\label{sec_phase_diagram}Phase diagram}

\subsection{\label{sec_threePhases}Three phases}

In this section, we show the phase diagram and explain it with respect to surface science.

As we have reported previously \cite{akutsu10}-\cite{akutsu14}, for $\epsilon_{\rm int}/\epsilon= -0.5$, there are two phase transition temperatures, $T_{f,1}$ and $T_{f,2}$. 
For $T>T_{f,1}$,  the surface free energy has the GMPT universal form (equation  (\ref{f-gmpt}), figures \ref{ecs} (c) and \ref{bgam} (c)).

For $T<T_{f,1}$, the surface gradient $\vec{p}$ jumps at $\eta_c^{(111)}$ of the (111) facet edge (point Q in figure  \ref{ecs} (b)).  
Here, the vicinal surface free energy, $f(\vec{p})$, for the homogeneous system, does not connect to the surface free energy  of the (111) surface.
Similarly, the surface tension also becomes disconnected around the (111) surface (figure  \ref{bgam} (b)).

For $T<T_{f,2}$, $\vec{p}$ jumps at $\eta_c^{(001)}$ of the (001) facet edge (point P in figure  \ref{ecs} (a)).
Then, the surface tension, $\gamma_{\rm surf}(\vec{p})$, near $|\vec{p}| \sim 0$ has an equilibrium value at only $|\vec{p}| =0$.
In our previous work \cite{akutsuJPCM11,akutsu12,akutsu14}, where $\epsilon_{\rm int}/\epsilon= -0.5$, the spinodal point was too close to the (001) surface to distinguish the spinodal point from the point of the (001) surface.

We call the phase where $T>T_{f,1}$ the {\it GMPT phase}, the phase where $T_{f,2}<T<T_{f,1}$ the {\it step droplet phase}, and the phase where $T<T_{f,2}$ the {\it step faceting phase}.
The phase diagram of the p-RSOS model with an area of $-2 \leq \epsilon_{\rm int}/\epsilon \leq 0$ and $0<\kBT/\epsilon \leq 2$ is examined by the PWFRG calculations.
The calculated phase diagram is shown in figure  \ref{phase_diagram}.

\subsection{\label{roughening}Roughening transition temperature}

The roughening transition temperature of the (001) surface, $\TR^{(001)}$, is determined so that the principal curvature, $\kappa$, may equal the universal value, $ 2/(\pi \kB \TR^{(001)})$, where $\kappa=\lim_{|\vec{\eta}| \rightarrow 0} \Delta p/\Delta |\vec{\eta}|$.
As seen in figure  \ref{phase_diagram}, the sticky character of the steps roughens the (001) surface.
Hence, the roughening transition temperature decreases as $\epsilon _{\rm int}$ decreases.
Around $\epsilon_{\rm int}/\epsilon =-2$, $T_{f,2}$  converges to $\TR^{(001)}$.

At temperatures $T>\TR^{(001)}$, the (001) surface is rough.
The step tension of a monostep, $\gamma_1(\phi)$, equals zero \cite{beijeren,jayaprakash} and the step stiffness of a monostep,  $\tilde{\gamma}_1(\phi)=\gamma_1(\phi)+\partial^2  \gamma_1(\phi)/\partial\phi^2 $, equals zero, where  $\phi$ is the tilting angle of the mean running direction of the steps relative to the $\langle 010 \rangle$ direction.
Hence, in the limit of $p \rightarrow 0$, the surface free energy has the form as follows:
\beq
f(\vec{p}) =f(0)+A (\phi) \frac{ |\vec{p}|^2}{d_1^2} 
 +C(\phi) \frac{ |\vec{p}|^4}{d_1^4}+{\cal O} (p^5). \label{fptr}
\eeq
where  $d_1$ (=1) is the height of a mono-step, $A(\phi)$ is the step coalescence factor, and $C(\phi)$ is the step-collision coefficient.
At $\TR$, $A(\phi)$ does not depend on $\phi$ and have the value of $ \pi \kB \TR^{(001)}/4$ \cite{jayaprakash}.

It should be noted that the roughening transition temperature of the (111) surface, $\TR^{(111)}$, is infinite in this model due to the RSOS restriction. 
Around (111) facet edge, the surface free energy shows the GMPT behavior.
Hence, we name the GMPT phase for $T\geq \TR^{(001)}$ GMPT-I and the GMPT phase for $T<\TR^{(001)}$ GMPT-II.

\subsection{\label{sec_step_faceting}Step faceting phase}

The disconnected shape of the surface tension caused macro-step formation as the two surface coexisted \cite{akutsu10}-\cite{akutsu14}.
All the steps condensed into one merged step at equilibrium.
The side surface  of the macro-step is a smooth (111) surface.
Examples of the macro-step formation are shown in figure  \ref{MC} (a) using the Monte Carlo method with the Metropolis algorithm.

Because the kink density on the side surface of a merged step is small,  the velocity of a merged step becomes much smaller than that of a single step.
Hence, the steps are pinned by the locally formed merged steps without adsorbates, impurities, or dislocations \cite{akutsu12,akutsu14}.

\subsection{\label{sec_step_droplet}Step droplet phase}

The step droplet phase occurs in the temperature range $T_{f,2}<T<T_{f,1}$ \cite{akutsuJPCM11,akutsu03}.
The roughening transition temperature, $\TR^{(001)}$, divides the step droplet phase into the step droplet-I phase ($T \geq \TR^{(001)}$) and the step droplet-II phase ($T< \TR^{(001)}$). 
We explain the step droplet-II phase first.

Near the (111) surface, where the step density is high, a macro-step of the (111) surface appears owing to the disconnected shape in the surface tension around the (111) surface.
The surface with a slope of $p^*_1$, which equals $p$ at point Q in figure  \ref{ecs} (b), coexists with the (111) surface instead of the (001) surface.
The $p^*_1$ is the end point of the solid line in figure  \ref{eta-p-0.9}  or in figure  \ref{bgam} (b).
Examples of the macro-step formation are shown in figure  \ref{MC} (b).

\begin{figure}
\centering
\includegraphics[width=6 cm,clip]{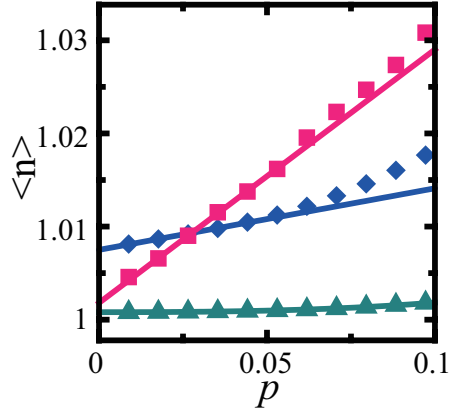}%
\caption{
Monte Carlo calculations of the average size of a locally merged step, $\langle n \rangle$ ($n$-mer).
Averaged over $4\times 10^8$ MCS.
Size: $160\sqrt{2} \times 160\sqrt{2}.$
$p=(N_{\rm step}/160)\sqrt{2}$.
Squares: $\epsilon_{\rm int}/\epsilon = -0.9$ and $\kBT/\epsilon=0.615$.
Triangles: $\epsilon_{\rm int}=0$ (original RSOS model) and $\kBT/\epsilon=0.615$.
Diamonds: $\epsilon_{\rm int}/\epsilon = -0.9$ and $\kBT/\epsilon=0.73$.
Red line: $ 1.0018 + 0.272 p$.
Green line: $1.0008+0.0294 p^2 + 0.738 p^3$.
Blue line: $ 1.0075 + 0.0657 p$.
}
\label{Navp0-0.9}
\end{figure}

For the surface slope $p<p^*_1$ (low step density), a homogeneous vicinal surface appears where the steps stick together locally.
In figure  \ref{Navp0-0.9}, we show the average size of a locally merged step calculated by the Monte Carlo method.
Here, $\langle n \rangle$ is defined by
\beq
\langle n \rangle = \frac{1}{\mathcal N'}\sum_{i,j} |h(i+1,j+1)-h(i,j)|
\eeq
where ${\mathcal N'}$ is the total number of uneven height differences.
Similar to the case  with $\epsilon_{\rm int}/\epsilon=-0.5$ in our previous paper \cite{akutsuJPCM11}, $\langle n \rangle$ increases linearly as $p$ increases in the limit of $p \rightarrow 0$ (figure  \ref{Navp0-0.9}).
However, $\langle n \rangle$ of the original RSOS model
($\epsilon_{\rm int}=0$) remains as 1 as $p$ increases near $p \sim 0$.

Owing to the $p$ dependence of $\langle n \rangle$, the expansion of $f(\vec{p})$ with respect to $|\vec{p}|$ is  rewritten as \cite{akutsuJPCM11} (\ref{p-expansion})
\beqa
f_{\rm eff}(\vec{p})&& =f(0)+\gamma_1 (\phi)  \frac{|\vec{p}|}{d_1}+A_{\rm eff} (\phi) \frac{ |\vec{p}|^2}{d_1^2} \nonumber \\
&&+B_{\rm eff} (\phi) \frac{ |\vec{p}|^3}{d_1^3} +C_{\rm eff} (\phi) \frac{ |\vec{p}|^4}{d_1^4}+{\cal O} (p^5), \label{fpeff0}
\eeqa
where  $A_{\rm eff}(\phi)$ is the effective step coalescence factor, $B_{\rm eff}(\phi)$ is the effective step interaction coefficient, and $C_{\rm eff}(\phi)$ is the effective step-collision coefficient.
Here, $\phi=0$ for the vicinal surface inclined in the $\langle 101 \rangle$ direction, and $\phi=\pi/4$ for the vicinal surface inclined in the $\langle 111 \rangle$ direction.
For simplicity, we assume $\phi=\pi/4$ and abbreviate  $\phi$ in equation  (\ref{fpeff0}), hereafter.

The special feature of equation  (\ref{fpeff0}) is the quadratic term with respect to $|\vec{p}|$.
The quadratic term is generated from the change in step tension of the $n$-merged step  \cite{akutsuJPCM11} (\ref{p-expansion}).

For the step droplet-II phase, where (001) surface is rough, the effective surface free energy has the form in the limit of $p \rightarrow 0$ as
\beqa
f_{\rm eff}(\vec{p})&& =f(0)+A_{\rm eff} |\vec{p}|^2 
+B_{\rm eff} |\vec{p}|^3 +C_{\rm eff}   |\vec{p}|^4
\nonumber \\
&&
+{\cal O} (p^5). \label{fpefftr}
\eeqa
Since the surface tension has a disconnected shape around the (111) surface, the surface with slope $p$ with $p_1^*<p<\sqrt{2}$ is realized as the coexistence of (111) surface and the surface with the slope $p_1^*$.
The image of the two surface coexistence is similar to figure  \ref{MC} (b).
For the surface slope $p<p^*_1$ (low step density), a homogeneous vicinal surface appears.

\section{\label{sec_1DQP}Exclusive sphere in 1D bosons}

\subsection{\label{sec_1dboson111}Surface free energy of a vicinal surface inclined in the $\langle 111 \rangle$ direction}

In this section, we explain the relationship between the original RSOS model ($\epsilon_{\rm int}=0$) and the 1D quantum particle systems (figure \ref{vicinalTSK}).

The 1D FF behaviour on a vicinal surface is well established and is the foundation of the GMPT universal behaviour (\ref{sec_correspond}) \cite{gmpt},\cite{izuyama}-\cite{yamamoto97}.
We summarize the correspondence between the surface system and 1D spinless quantum particle system in Table \ref{correspond}. 
For the quantum particle system in a 1D lattice, the ground state energy is expressed in the same form as equation  (\ref{ground_eFF}).

The key concept for mapping the surface steps to the 1D FF system is the  impenetrability between quantum particles. 
On a vicinal surface below the roughening transition temperature for a terrace surface, there are almost no {\it overhung} structures.
Hence,  the adjacent steps do not cross each other, and this is considered to be impenetrability \cite{tonks} in a 1D quantum particle system \cite{jayaprakash84-2,akutsu88,yamamoto89,yamamoto94,yamamoto95,yamamoto96,yamamoto97}.

\begin{table}
\caption{\label{correspond}Correspondence}
\footnotesize
\centering
\begin{tabular}{@{}ll}
\br
Surface system & 1D Bose system\\
\mr
(001) surface & Vacuum \\
(111) surface & Close packed solid\\
& or the vacuum for a negative step\\
Step (down) & Particle\\
Step (up) & Antiparticle\\
Negative step & Vacancy\\
Mean running &  Imaginary time\\
direction 	of steps & \\
&\\
Step density: & Density of quantum \\
$\rho=|\vec{p}|/d_1$$^a$    &  particles: $\rho^{\rm QP}$\\
Andreev field: $\pm d_1 |\vec{\eta}| $& Chemical potential of \\
	&  quantum particles: $\mu$  \\
Step tension: $\gamma$ &  $\mu_0=\mu|_{\rho^{\rm QP} \rightarrow 0}$\\
Surface free energy& Ground state energy \\
 (density): $f(\vec{p})$ &  at $T^{\rm QP}=0$: $E(\rho^{\rm QP})$\\
\br
\end{tabular}\\
$^{a}$ $d_1$ is the height of a step.
\end{table}

If we consider a coarse-grained vicinal surface, the Hamiltonian for the transfer matrix is mapped to the continuous model of quantum particles in one dimension \cite{akutsu88}.
Even in this case, the impenetrability between quantum particles is crucial.  
Girardeau  \cite{girardeau} showed that the wave function at absolute zero ($T^{\rm QP}=0$) of 1D impenetrable spinless bosons is equivalent to that of  1D fermions.



\begin{table*}
\caption{\label{parameter}Calculated parameters for $\epsilon_{\rm int}=0$ and $p \sim 0$. }
\footnotesize
\centering
\begin{tabular}{@{}llllllll}
\br
$\phi$ & $\kBT/\epsilon$ & $\gamma/(\sqrt{2}\kBT)$& $3\sqrt{2}B/(\kBT)$ & $z$ &  $4C_4/(\kBT)$ &   $\sigma$ & $c$  \\
\mr
 $\pi/4$  & 0.2 & 4.307 & 4.88 &4.2  & $-2.3$ & 0.07 & 0.5  \\
 $\pi/4$  & 0.4 & 1.800 & 5.22 &4.0  & $-4.2$ & 0.11 & 0.7  \\
 $\pi/4$  & 0.6 & 0.939 & 6.36 &4.0  & $-4.3$ & 0.15 & 1.5  \\ 
$\pi/4$  & 0.8 & 0.486 & 9.83 &4.1  & $-8.7$ & 0.3 & 2  \\ 
 $\pi/4$  & 1.0 & 0.212 & 22 &4.1 & $-58$ & 1.6 & 2  \\
\br
\end{tabular}
\end{table*}

The vicinal surface inclined in the $\langle 111 \rangle$ direction (figure  \ref{vicinal}), however, cannot be mapped to 1D fermion systems because a lattice site can be occupied by two steps.
Besides, more than two steps cannot occupy one site at a time due to the RSOS restriction.
Because of this ``quasi-impenetrability'', the vicinal surface inclined in the $\langle 111 \rangle$ direction should be mapped to a 1D spinless boson system (\ref{sec_boson}).

\begin{figure}
\centering
\includegraphics[width=6 cm,clip]{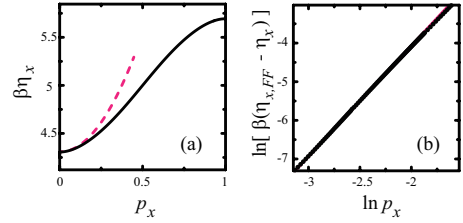}%
\caption{
$p_x$ dependence of $\beta \eta_x$ for a vicinal surface inclined in the (111) direction ($\phi=\pi/4$), where $\beta= 1/\kBT$ and $p_y=p_x$, calculated with the RSOS model.
$\epsilon_{\rm int}=0$.
$\kBT/\epsilon=0.2$.
(a) Solid line: PWFRG calculations.
Broken line: $\beta \eta_{x,FF}=  \beta \gamma(\pi/4)/(\sqrt{2}d) + 3\sqrt{2} \beta B(\pi/4)p^2/d^3$, where $d=a=1$.
(b) 
Closed circles: $\ln [ \beta (\eta_{x,FF}-\eta_x)] $ for the PWFRG calculations.
Solid line: $\ln [ \beta (\eta_{FF}-\eta)] = 2.9 \ln p + 1.67$.
}
\label{eta-p}
\end{figure}

We analyzed the $p_x(\vec{\eta}) $ values calculated with PWFRG (figure  \ref{eta-p} (a))  by using the least-squares method  in several ways (figure  \ref{eta-p}).
From the analysis, the form of $f(\vec{p})$ is
\beq
f(\vec{p})=f(0) +\gamma + B p^3 
+C_4p^4 +\mathcal{O}(p^5), \label{eqfit}
\eeq
where $p=|\vec{p}|$ and $d_1=1$.
The parameters obtained for several temperatures in the RSOS system are listed in Table \ref{parameter}.

Because there is a fourth-order term with respect to $p$ in equation  (\ref{eqfit}), the surface free energy of the vicinal surface inclined in the (111) direction does not agree with that of a pure FF system.
Comparing the ground state energy, equation  (\ref{ground_eBose}), with the surface free energy, equation  (\ref{eqfit}), we can obtain the values of $\sigma$ and $c$ \cite{okunishi99} which are listed in Table \ref{parameter}.
$\sigma$ and $c$ increase as the temperature of the RSOS system increases.

In our previous work \cite{yamamoto89},  we derived equation  (\ref{ground_eFF}) exactly for the vicinal surface inclined in the $\langle 111 \rangle$ direction with the RSOS model.
In that work, strict impenetrability between steps was required; more than one step could not occupy one site at a time.
We regarded a step on the surface as a trajectory of a random walker in a 1D lattice.
The random walker traced the path on the 2D skewed lattice.
The operators for the random walkers required the anti-commutator relation to satisfy the impenetrability.
The partition function of the random walkers are transferred to that of the 1D FF system by applying the Jordan-Wigner transformation \cite{jordan}.

Therefore, we can conclude that the meeting of neighbouring steps at a site, which corresponds to the collision of neighbouring quantum particles on the 1D lattice, makes  the $C_4(\pi/4)\rho^4$ term the surface free energy, and thus the ground state energy of 1D bosons.

\subsection{\label{sec_1dboson101}Vicinal surface inclined in the $\langle 101 \rangle$ direction}

The step faceting phase and the step droplet phase do not appear in the vicinal surface inclined in the $\langle 10 1 \rangle$ direction.
To understand why the vicinal surface inclined in the $\langle 10 1 \rangle$ direction does not show the step faceting phase, in this subsection we confirm that the surface free energy of the vicinal surface inclined in the $\langle 101 \rangle$ direction behaves like the pure FF, where $f(\rho)$ has no $C_4\rho^4$ term.

\begin{figure}
\centering
\includegraphics[width=6 cm,clip]{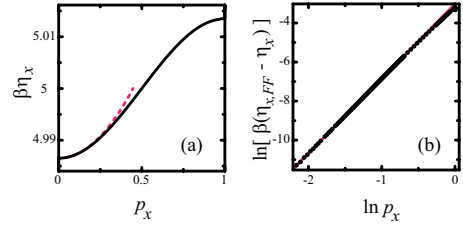}%
\caption{
$p_x$ dependence of $\beta \eta_x$ for a vicinal surface inclined in the (101) direction ($\phi=0$), where $\beta= 1/\kBT$ and $p_y=0$, calculated with the RSOS model.
$\epsilon_{\rm int}=0$,
$\kBT/\epsilon=0.2$.
(a) Solid line: PWFRG calculations.
Broken line: $\beta \eta_{x,{\rm FF}}= \beta \gamma(0)/d + 3\beta B(0)p^2/d^3$, where $d=a=1$.
(b) 
Closed circles: $\beta (\eta_{x,{\rm FF}}-\eta_x)$ for the PWFRG calculations.
Solid line: $\beta (\eta_{x,{\rm FF}}-\eta) = 3.8 \ln p - 3.0$.
}
\label{eta-p101}
\end{figure}

\begin{table}
\caption{\label{parameterFF}Calculated parameters for  $\epsilon_{\rm int}=0$ and $p \sim 0$. }
\footnotesize
\centering
\begin{tabular}{@{}llllll}
\br
$\phi$ & $\kBT/\epsilon$ & $\gamma/(\kBT)$ & $3B/(\kBT)$ & $z$ &  $5C_5/(\kBT)$ \\
\mr
 $0$  & 0.2 & 4.98652 & 0.0673 &4.8  & -0.0611\\
 $0$  & 0.3 & 3.26196 & 0.376 &4.8    & -0.611 \\
 $0$  & 0.4 & 2.3356 & 0.942 &5.0    & -2.32  \\
 $0$  & 0.6 & 1.287 & 2.9 &4.9    & -19\\ 
\br 
\end{tabular}
\end{table}

By using the same method as for the vicinal surface inclined in the $\langle 111 \rangle$ direction, we find that the next order to $\rho^3$ with respect to $\rho$ is $\rho^5$ (figure  \ref{eta-p101}).
Finally, we obtained the form for the surface free energy as
\beqa
f(\vec{p})&=&f(0) +\gamma(0)p + B(0) p^3 +C_5(0)p^5
\nonumber \\
&&
 +\mathcal{O}(p^6), \label{eqfitFF}
\eeqa
where $\phi=0$, $d_1=1$, and $C_5(0)$ represents the coefficient of the $p^5$ term.
The numerical values obtained by the analysis to derive equation  (\ref{eqfitFF}) are listed in Table \ref{parameterFF}. Comparing equation  (\ref{eqfitFF}) with the ground state energy of FFs equation  (\ref{ground_eFF}) shows that the vicinal surface inclined in the $\langle 101 \rangle$ direction behaves like a typical pure FF system.

\subsection{\label{sec_exclusive_sphere} Effective exclusive sphere}

If we consider a coarse-grained vicinal surface in the original RSOS model,  the transfer matrix Hamiltonian of the vicinal surface inclined in the $\langle 101 \rangle$ direction is equation  (\ref{hamilFF}), whereas the Hamiltonian of the vicinal surface inclined in the $\langle 111 \rangle$ direction is equation  (\ref{hamilLieb}) for small $\rho$.
In this subsection, we discuss what makes the transfer Hamiltonian in the continuous system different.

Here, we introduce the radius of the effective exclusive sphere, $r_{\rm ex}$, and the effective force range, $r_f$, for the point-contact-type step-step repulsion in the original RSOS model. 
If we push the steps on the vicinal surfaces together, we obtain a (111) surface for the vicinal surface inclined in the $\langle 111 \rangle$ direction, and a ($101$) surface for the vicinal surface inclined in the $\langle 101 \rangle$ direction.
The maximum step density for the (111) surface, $\rho_{\rm max}^{(111)}$, equals $\sqrt{2}$ ($d=a=1$) and the maximum step density for the ($101$) surface, $\rho_{\rm max}^{\langle 101 \rangle}$ equals 1.
Hence, we define the effective exclusive sphere for the (111) surface and for the (101) surface with sphere radii of $r_{\rm ex}^{\langle 111 \rangle}=1/(2 \sqrt{2})$ and $r_{\rm ex}^{\langle 101 \rangle}=1/2$, respectively.

However, if we regard the meeting of the step as the point-contact-type step-step repulsion, the effective force range, $r_f$, is $1/\sqrt{2}$.
For the coarse-grained vicinal surface inclined in the $\langle 111 \rangle$ direction, $2r_{\rm ex}^{\langle 111 \rangle}=r_f$.
Considering that $\lim_{L \rightarrow \infty} r_{\rm ex}^{\langle 111 \rangle}/L \rightarrow 0$, the transfer matrix Hamiltonian converges to equation  (\ref{hamilLieb}) in the limit of $\rho \rightarrow 0$.

For the vicinal surface inclined in the $\langle 101 \rangle$ direction, $2r_{\rm ex}^{\langle 101 \rangle}$ is larger than $r_f$.
The strict impenetrability is satisfied, and the point-contact-type step-step repulsion does not work.
Hence, the transfer matrix Hamiltonian converges to equation  (\ref{hamilFF}) in the limit of $\rho \rightarrow 0$. 

Therefore, the magnitude relationship between $2r_{\rm ex}$ and $r_f$ affects the form of the Hamiltonian for the transfer matrix in the continuous 1D system.

\section{\label{sec_1Dattractive_bosons}Phase separation lines}

\subsection{Bose solid and Bose fluid \label{sec_1DBose_solid}}

In this subsection, we interpret the phases in the phase  diagram (figure  \ref{phase_diagram}) from the perspective of the 1D spinless attractive bosons with quasi-impenetrability.

For the attractive ($\epsilon_{\rm int}<0$) p-RSOS model, the step faceting phase occurs at low temperatures ($T<T_{f,2}$, figure  \ref{phase_diagram}) for the vicinal surface inclined in the $(111)$ direction.
At a slope $0<p<\sqrt{2}$, the surface consists of the (001) surface (terrace) and a (111) surface (faceted macro-step).
This structure is mapped to  the vacuum and a Bose solid (figure  \ref{MC} 8a)), respectively (Table \ref{correspond}).

The Bose solid state, $\Psi(x)$, where $x$ is the site of the left edge of the Bose solid, is a bound state where all the quantum particles stick together, and is a condensed state for the spinless bosons.
However, the condensate is not the same as the Bose-Einstein condensate because of the quasi-impenetrability.

The effective exclusive sphere is crucial to forming the Bose solid with a finite volume.
If we consider the system in equation  (\ref{hamilLieb}) \cite{lieb} with $c<0$, all the bosons collapse to a point.
However, for the coarse-grained vicinal surface inclined in the $\langle 111 \rangle$ direction, the density has a maximum value of $\rho_{\rm max}^{\rm QP} = \sqrt{2}$ owing to the effective exclusive sphere.
Hence, the Bose solid occupies a finite volume.


The step droplet phase is interpreted as a mixture of a 1D Bose solid and 1D Bose fluid.
A 1D Bose solid (faceted step) appears when $\rho$ is sufficiently high (figure  \ref{MC} (b)).
The Bose fluid state considered here is a mixture of boson $n$-mers.
Because the bound state of a boson $n$-mer is not the eigenstate of the system in the Bose fluid phase, the boson $n$-mer has a finite lifetime.
The bosons change from $n$-mer to $(n+m)$-mer ($m= \pm 1$, $\pm 2$, $\cdots$) over time. 

The 1D Bose fluid at the temperatures $T>\TR^{(001)}$ is different from the one at the temperatures $T_{f,2}<T<\TR^{(001)}$.
We call the 1D Bose fluid for $T>\TR^{(001)}$ 1D Bose fluid-I, and the one for $T_{f,2}<T<\TR^{(001)}$ 1D Bose fluid-II, hereafter. 
In the limit of $\rho^{QP} \rightarrow 0$, the chemical potential of the 1D Bose fluid-I, $\mu_0$ ($\mu_0=\gamma_1=\eta_c^{(001)}$), is zero (gapless);  however, $\mu_0$ of the 1D Bose fluid-II is finite.
It should be noted that, in the limit of $\rho^{QP} \rightarrow \sqrt{2}$, the chemical potential of the 1D boson for the ``negative step'' (\S \ref{sec_negative_step}) is finite for $T>\TR^{(001)}$.

We consider that the 1D Bose fluids-II in the step droplet phase-II and in the GMPT phase-II are the same.
The difference between the two phases may be $A_{\rm eff}$. In the GMPT phase, $A_{\rm eff}=0$  (figure  \ref{eta-p-0.9}) in the low density limit.
However, $A_{\rm eff}>0$ near $T_{f,2}$ in the step droplet phase  (figure  \ref{eta-p-0.9}).    
By using the step quantities, $A_{\rm eff}$ is expressed as (\ref{p-expansion})
\beq
A_{\rm eff}   =n_0^{(1)}   \gamma_1^{(1)}   /d_1, \label{eqAeff0} 
\eeq
where $n_0^{(1)}=\partial \langle n \rangle /\partial p |_{p \rightarrow +0}$, $p=|\vec{p}|$, $\gamma_1^{(1)}=\partial (\gamma_n /n)/\partial n|_{n \rightarrow +1} $,   $\gamma_n$ is the step tension of an $n$-merged step, and $d_1=1$.

In the step droplet phase-II, $n_0^{(1)}>0$, as in figure  \ref{Navp0-0.9}.
Even in the GMPT phase-II, we obtained a small but positive $n_0^{(1)}$  in the least-squares fitting to the values obtained by the Monte Carlo method  (figure  \ref{Navp0-0.9}), which contrasts with $n_0^{(1)}=0$ in the original RSOS model.
Thus, on average, boson $n$-mers with a finite lifetime survive in the low-density limit  in the Bose fluid-II state.
Over time, the boson $n$-mers evaporate and are formed because of quantum fluctuations.

From equation  (\ref{eqAeff0}), positive $n_0^{(1)}$ leads to positive $\gamma_1^{(1)}$  for the step droplet phase-II near $T_{f,2}$, and $\gamma_1^{(1)}=0$ for the GMPT phase-II \cite{akutsuJPCM11}.
As temperature increases from near $T_{f,2}$, $A_{\rm eff}$ and $n_0^{(1)}$ decrease.
Hence, at high temperatures, such as near $T_{f,1}$, it is difficult to detect finite $A_{\rm eff}$ in the PWFRG calculations.
The border between $\gamma_1^{(1)}>0$  and $\gamma_1^{(1)}=0$ is so subtle that we cannot determine the lowest temperature numerically with $\gamma_1^{(1)}=0$ for fixed $\epsilon_{\rm int}$.

\subsection{Solidification at low densities \label{psl_Tf2}}

\subsubsection{Lower phase separation line}

The lower PSL is the line $T_{f,2}^{\rm PSL}$, which is obtained from the condition $\gamma_1= \lim_{n \rightarrow \infty} \gamma_n/n$.
At the temperatures $T> T_{f,2}^{\rm PSL}$,  $\gamma_1< \lim_{n \rightarrow \infty} \gamma_n/n$, whereas at $T< T_{f,2}^{\rm PSL}$, $\gamma_1> \lim_{n \rightarrow \infty} \gamma_n/n$.
By using the imaginary path-weight random walk (IPW) method \cite{akutsu90} on the 2D square Ising model, we obtain the following equation \cite{akutsuJPCM11}.
\beq
\frac{\cosh^2 (\epsilon/\kBT_{f,2}^{\rm PSL})}{\sinh (\epsilon/\kBT_{f,2}^{\rm PSL})}  = 2 \cosh \left (\frac{2\epsilon+ \epsilon_{\rm int}}{2 \kBT_{f,2}^{\rm PSL}}\right ).
\label{kBTf2}
\eeq
The calculated PSL curve (equation  (\ref{kBTf2})) is shown in figure  \ref{phase_diagram} as a solid line.

Note that $\mu_{c}=\eta_c^{(111)}$ (figure  \ref{ecs}) corresponds to the  chemical potential that causes the  condensation of bosons.
The condition for deriving equation  (\ref{kBTf2}) is expressed as $\mu_0=\mu_c$.

The agreement between equation  (\ref{kBTf2}) and PWFRG-calculated values decreases at about $\epsilon_{\rm int}/\epsilon < -1.5$.
This is because $T_{f,2}$ approaches the roughening transition temperature of the (001) surface, $\TR^{(001)}$.
Near $\TR^{(001)}$, multi-level islands and negative islands are formed frequently.
Various islands are visible on the terrace at $\kBT/\epsilon = -1.3$ in figure  \ref{MC}.
$\gamma_1$ in equation  (\ref{kBTf2}) is estimated by the IPW method based on the 2D Ising model, which is the model used for two-level problems.
Therefore, the approximation becomes worse near the roughening transition temperature.

\subsubsection{\label{sec_ecs}Shape exponent near $T_{f,2}$}


The form of the surface free energy equation  (\ref{fpeff0}) affects the shape of the crystal droplet at equilibrium (figures \ref{ecs} and \ref{eta-p-0.9}).

Here, we define the shape exponent of the normal direction, $\theta_n$, as $z(x_c,y_c)-z(x,y)\propto |\vec{r}-\vec{r}_c|^{\theta_n}$ in the limit of $\vec{r} \rightarrow \vec{r}_c$, where $\vec{r}_c=(x_c,y_c)$ represents a point on a facet edge (e.g., point P in figure  \ref{ecs}), and $\vec{r}$ runs along the principal axis normal to the facet edge.
In the GMPT phase, $\theta_n=3/2$ in the limit of $p \rightarrow 0$.

 In contrast, in the step droplet-II phase, $\theta_n=2$ \cite{akutsuJPCM11}  owing to the positive step coalescence factor, $A_{\rm eff}$.
We can see this in figure  \ref{eta-p-0.9} in the limit of $p \rightarrow 0$. 
For the original RSOS model, the result of $n_0^{(1)}=0$ (figure  \ref{Navp0-0.9}) leads to $A_{\rm eff}=0$ because of the equation  (\ref{eqAeff0}) (\ref{p-expansion}).
As temperature increases, $A_{\rm eff}$ decreases rapidly.
We cannot determine the value of $A_{\rm eff}$ at higher temperatures by analysing the values obtained by PWFRG calculations.

The shape exponent, $\theta_n$, can be  translated to a 1D quantum particle system as
\beq
E(-\mu_0)-E(-\mu) \propto |\mu-\mu_0|^{\theta_n}.
\eeq

\subsection{Melting at high densities \label{critical_curve}}

\subsubsection{Negative step\label{sec_negative_step}}

\begin{figure}
\centering
\includegraphics[width=8 cm,clip]{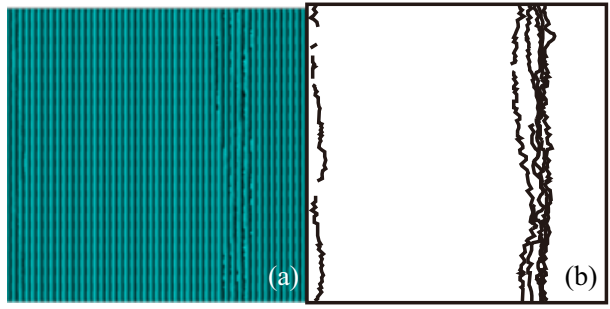}%
\caption{
(a) Snap shot of the top view of a vicinal surface for Monte Carlo calculations.
$\kBT/\epsilon=0.709$.   $\epsilon_{\rm int}/\epsilon = -0.9$.
Size: $240\sqrt{2} \times 240\sqrt{2}.$
$N_{\rm step}=474$.
$4\times 10^8$ MCS.
(b) Negative steps in (a).
}
\label{mcT709}
\end{figure}

The upper PSL is line $T_{f,1}^{\rm PSL}$, which is more complex than the lower PSL.
From the definition, the anomaly in surface tension occurs near $p= \sqrt{2}$.
At $\mu>\mu_c=\eta_c^{(111)}$, bosons solidify to form a (111) surface (figure  \ref{ecs}).
The vicinal surface near the (111) surface corresponds to a mixture of the Bose solid and vacancies (figure  \ref{mcT709}), where $\mu$ is close to $\mu_c$ and $\mu<\mu_c$.
Then, studying the vicinal surface near the (111) surface in the present model corresponds to studying the melting of a Bose solid of 1D spinless quasi-impenetrable bosons.
For $T>T_{f,1}$, the melting of Bose solid occurs as $\mu$ decreases, satisfying GMPT behaviour in the GMPT-I phase and the GMPT-II phase.
For $T<T_{f,1}$, the Bose solid melts as a first-order phase transition with respect to $\mu$  in the step droplet-I phase and the step droplet-II phase.
In this subsection, we examine the melting of Bose solid further.

\begin{figure}
\centering
\includegraphics[width=7 cm,clip]{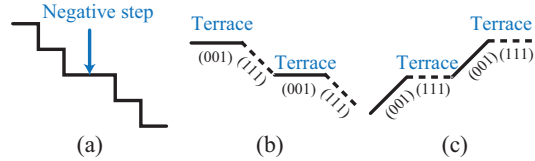}%
\caption{
(a) Example of a negative step.
(b) Side view of faceted steps. 
(c) View of (b) rotated by $45^{\circ}$. 
The side surfaces of the steps in (b) become terraces.
} 
\label{negativeStep}
\end{figure}

To study the vicinal surface with a slope near $\sqrt{2}$, we rotate the surface so that the surface slope of the (111) surface is zero.
Then, $(p, \eta)$ is mapped to $(p', \eta')$.
We introduce $q=-p'$ as the slope from the (111) facet such that $q= (\sqrt{2}-p)/(1+\sqrt{2}p)$.
We restrict ourselves to treating the case of $p_x=p_y$.

For the vicinal surface near a (111) facet, the (111) surface forms a terrace, and a (001) surface forms the side surface of a macro-step (figure  \ref{negativeStep} (b) and (c)).
We call the step with the (001) side surface a ``negative step'', hereafter (figure  \ref{negativeStep} (a)).
The negative steps are mapped to vacancies in a Bose solid, and they behave like 1D quasi-impenetrable bosons owing to the RSOS restriction (figure  \ref{mcT709} (b)).

To demonstrate that the negative steps behave like 1D quasi-impenetrable bosons, we calculate surface quantities for the original RSOS model ($\epsilon_{\rm int}=0$) around the (111) surface as follows.
Repeating the process explained in \S \ref{sec_1dboson111} with $\epsilon_{\rm int}=0$, we obtained the surface free energy for negative steps as
\beq
f(\vec{q})=f(0) +\gamma_q q + B_q q^3 
+C_{q,4}q^4 +\mathcal{O}(q^5), \label{eqfitq}
\eeq
where  $d_1=1$, $\gamma_q$ is the step tension for a negative step, $B_q$ is the step interaction coefficient, and $C_{q,4}$ is the step colliding coefficient.
The values obtained for $\gamma_q$, $B_q$, and $C_{q,4}$ are listed in Table \ref{parameterp=1}. Interestingly, $B_{q,1} /\kBT$ and $C_{q,4}/\kBT$ are almost constant as the temperature changes.

\begin{table*}
\caption{\label{parameterp=1}Calculated parameters for a negative step, \\$\epsilon_{\rm int}=0$ and $q \sim 0$ ($p \sim \sqrt{2}$). }
\footnotesize
\centering
\begin{tabular}{@{}llllllll}
\br
$\phi$ & $\kBT/\epsilon$ & $\gamma_q/(\kBT)$ & $3B_q/(\kBT)$ & $z$&  $4C_{q,4}/(\kBT)$ &   $\sigma_q$ & $c_q$  \\
&&&&&$\times 10^2$ &&\\
\mr
 $\pi/4$  & 0.2 & 1.698 & 48.1 &4.1  & $-1.97$ & 1.0 & 1.3  \\
  $\pi/4$  & 1.0 & 1.698 & 47.7 &4.2 & $-2.05$ & 4.8 & 6.0  \\
$\pi/4$  & 2.0 & 1.699 & 48.8 &4.1 & $-2.41$ & 9.9 & 11  \\
\br
\end{tabular}
\end{table*}

\begin{figure}
\centering
\includegraphics[width=6 cm,clip]{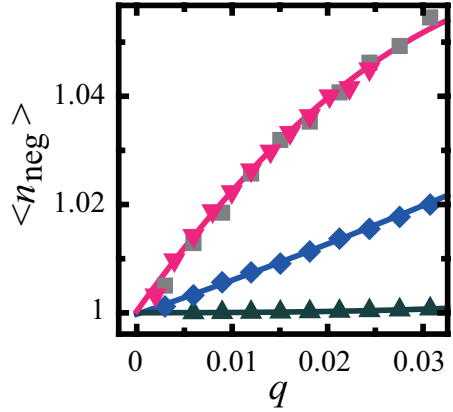}%
\caption{
Monte Carlo calculations of the averaged size of a  merged negative step, where $\langle n_{\rm neg} \rangle = \langle n \rangle (\sqrt{2}-p)/p $, and $q=(\sqrt{2}-p)/(1+\sqrt{2} p)$. 
Squares: standard $\kBT/\epsilon=0.709$, $\epsilon_{\rm int}/\epsilon = -0.9$, size $=160\sqrt{2} \times 160\sqrt{2}$, and averaged over $4\times 10^8$ MCS. 
Inverted triangles:  size $=240\sqrt{2} \times 240\sqrt{2}$, and averaged over $8\times 10^8$ MCS.
Diamonds: $\kBT/\epsilon=0.8$, averaged over $2\times 10^8$ MCS.
Triangles: $\epsilon_{\rm int}=0$ (original RSOS model), averaged over $1\times 10^8$ MCS.
Red line: $1.000+2.48 q - 25.3 q^2$.
Blue line: $1.000+0.609 q + 2.01q^2$.
Green line: $1.000+0.337 q^2 + 13.4 q^3$.
}
\label{navT709}
\end{figure}

Similar to $n$ near $p=0$, we consider the size of the merged negative step $n_{neg}$ near $p=\sqrt{2}$.
The relationships are
$
\langle n \rangle =(N_{\rm step})/({\langle n_{\rm step} \rangle })$, 
$\langle n_{neg}  \rangle =(2N-N_{\rm step})/({\langle \bar{n}_{\rm step}  \rangle})$, and 
$\bar{n}_{\rm step}=n_{\rm step} $,
where $N_{\rm step}$ is the total number of elementary steps,  $n_{\rm step}$ is the number of merged steps, and $\bar{n}_{\rm step}$ is the number of merged negative steps.
Hence, we have
\beq
\langle n_{\rm neg}\rangle = \langle n\rangle \frac{\sqrt{2}-p}{p}
\eeq
for $p=(N_{\rm step} d_1)/(\sqrt{2}a N)$ with $d_1=a=1$.
The $q$ dependence of $\langle n_{\rm neg}\rangle$ calculated by a Monte Carlo method in figure  \ref{navT709} shows that $\langle n_{\rm neg}\rangle \sim 1+ c' q + \cdots$ for the p-RSOS model, whereas  $\langle n_{\rm neg}\rangle \sim 1+ c'' q^2 + \cdots$ for the original RSOS model.
Thus, the $q$ dependence of $\langle n_{\rm neg}\rangle$ of the p-RSOS model is  different from that of the original RSOS model.

Based on the argument in \S \ref{sec_1dboson111} and  \ref{p-expansion} for the negative steps, the surface free energy  $f(\vec{q})$ with $q_x=q_y$ around the (111) facet is expressed as
\beqa
f_{\rm eff}(\vec{q})&& \equiv f_{\langle n_{\rm neg} \rangle}(\vec{q})=f(0)+\gamma_{q,1}   q
+A_{q, {\rm eff}}   q^2 \nonumber \\
&&+B_{q, {\rm eff}}   q^3 +C_{q, {\rm eff}}  q^4+  
{\cal O} (q^5), \label{fpeff_neg}
\eeqa
where $\gamma_{q,1}$ is the step tension of a negative monostep.

Note that the roughening transition temperature of the (111) surface is infinite in the RSOS model.
The negative steps in the step droplet-I (GMPT-I) phase behave similar to the ones in the step droplet-II (GMPT-II) phase.

\subsubsection{\label{sec_upperPSL}Upper phase separation line}

\begin{figure}
\centering
\includegraphics[width=8 cm,clip]{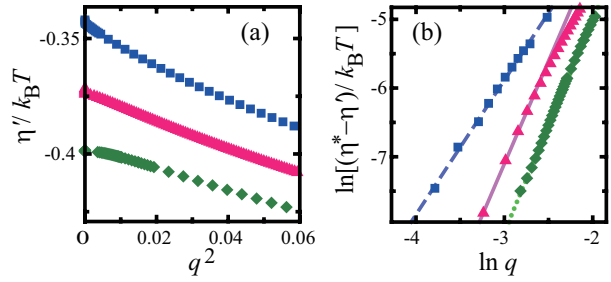}%
\caption{
(a) $q^2-\eta '$ curve around the (111) facet obtained by PWFRG calculations.
$\epsilon_{\rm int}/\epsilon = -0.9$.
Closed diamonds: $\kBT/\epsilon=0.7$.
Closed triangles: $\kBT/\epsilon=0.709$.
Closed squares: $\kBT/\epsilon=0.72$.
(b) Data in (a) replotted on logarithmic axes. 
Lines are the best-fit lines obtained by the least-squares method to find the exponent $z$.
The values of $z$ are listed in Table \ref{tf1psl}.
}
\label{eta-pT709}
\end{figure}

\begin{table}
\caption{\label{tf1psl}The comparison of $\eta_c^{(111)}/\kBT$ and $\eta^*/\kBT$}
\footnotesize
\centering
\begin{tabular}{@{}llllllll}
\br
$\kBT/\epsilon$ & $\eta_c^{(111)}/\kBT$ & $\eta^*/\kBT$ & z$^{a}$\\
\mr
0.72 & $-0.3432 \pm 0.0002$ & $-0.34338 \pm 0.00006$ & 2.96 \\
0.709 & $-0.3748 \pm 0.0002$ & $-0.37414 \pm 0.00006$& 3.99 \\
0.7 & $-0.39910 \pm 0.00008$ & $-0.39863 \pm 0.00006$ & 4.10 \\
\br
\end{tabular}\\
$^{a}$$(\eta^* - \eta')/\kBT \propto q^{z-1}$.
\end{table}

In figure  \ref{eta-pT709} (a), we show the relationship between $\eta'$ and $q^2$ calculated by the PWFRG method for $\epsilon_{\rm int}/\epsilon = - 0.9$.
The facet edge of the (111) facet, $\eta_c^{(111)}$, is shown in Table \ref{tf1psl}.
At $\kBT/\epsilon=0.72$ in figure  \ref{eta-pT709} (a), GMPT behaviour occurs in the limit $q \rightarrow 0$ ($p \rightarrow \sqrt{2}$).
In contrast, at $\kBT/\epsilon=0.709$ and $\kBT/\epsilon=0.7$, the curves are convex near $q=0$.
We plot $\ln[\eta^*-\eta']$ {\it vs.} $\ln q$  in figure  \ref{eta-pT709} (b), where $\eta^*$ is a fitted facet edge.
For $\kBT/\epsilon=0.72$ in Table \ref{tf1psl}, $\eta^*$ agrees with $\eta_c$ within the error.
However, for $\kBT/\epsilon=0.709$ and  $\kBT/\epsilon=0.7$, we have $\eta_c < \eta^*$,  which means that $q$ jumps from 0 to a finite value near $q=0$. 
Therefore, $T_{f,1}^{PSL}=0.709\pm 0.004$ for $\epsilon_{\rm int}/\epsilon=-0.9$.
$T_{f,1}^{PSL}$ values for other $\epsilon_{\rm int}$ are obtained in the same way.

\subsubsection{\label{shape_exponent2}Shape exponent on $T_{f,1}$}

Exponent $z$, which is defined by $(\eta^* - \eta')/\kBT \propto q^{z-1}$,  changes from 3 to 4 (Table \ref{tf1psl}) as the temperature decreases to cross $T_{f,1}^{PSL}$.
This means that, on only $T_{f,1}^{PSL}$, the shape exponent $\theta_n$ at the (111) facet edge equals 4/3.

In addition,  on only the PSL, this  reveals 
\beq
A_{q, {\rm eff}}=0,\quad B_{q, {\rm eff}} =0, \quad C_{q, {\rm eff}}>0. \label{eq_tf1}
\eeq
$B_{q, {\rm eff}}$ is positive at $T> T_{f,1}$; whereas it is negative at $T< T_{f,1}$.  
$C_{q, {\rm eff}}$ is positive at temperatures of $\kBT/\epsilon <0.715$, it decreases as temperature increases, reaches zero around $\kBT/\epsilon =0.715$, and becomes negative at higher temperatures.

Based on equations. (\ref{fpeff_neg}),  (\ref{eqAeff0}),  (\ref{eqBeff}), and (\ref{eqCeff}) for negative steps, for the conditions $A_{q, {\rm eff}}=0$ and $B_{q, {\rm eff}}=0$ we have the equations
\beqa
&&
\gamma_{q,1}^{(1)}  =0, \label{eq_gam1},\\
&&
\gamma_{q,1}^{(2)}   =  -2 B_{q,1}\left\{ n_{q,0}^{(1)}\right\} ^{-2}. \label{eq_gam2}
\eeqa
Because $B_{q,1}>0$, we then have $\gamma_{q,1}^{(2)}<0$ on  $T_{f,1}$.

Furthermore, by substituting  equations (\ref{eq_gam1}) and (\ref{eq_gam2}) into equation  (\ref{eqCeff}) for negative steps, we have
\beqa
C_{q, {\rm eff}}&& =\frac{n_{q,0}^{(2)} }{n_{q,0}^{(1)} }\{ B_{q, {\rm eff}} -B_{q,1} \}
 \nonumber\\
&&
- B_{q,1} n_{q,0}^{(1)}   \left(  4  +  \frac{\tilde{\gamma}_{q,1}^{(1)} }{\tilde{\gamma}_{q,1}  }\   \right) 
+ C_{q,4}. \label{eqCeff3}
\eeqa
equation  (\ref{eqCeff3}) shows how $C_{q, {\rm eff}}$ relates to $B_{q, {\rm eff}}$.
Because $n_{q,0}^{(2)}<0$ (figure  \ref{navT709}) and $B_{q, {\rm eff}} \sim 0$ near $T_{f,1}$, the first term on the right-hand side of equation  (\ref{eqCeff3}) takes a positive value near $T_{f,1}$.
Therefore, $C_{q, {\rm eff}}$ can be  positive near $T_{f,1}$.

Next, we explain why the shape exponent on $T_{f,1}$ is different from the shape exponent near $T_{f,2}$.
The expression for the effective surface free energy, equation  (\ref{fpeff_neg}), is obtained in a similar way in  \ref{p-expansion}.
The difference originates from the difference in the roughness on the side surface of the merged step.
As seen in figure  \ref{mcT709}, the assembled negative steps are diffuse. 
This affects the $n$ dependence of $\gamma_n$.

\section{\label{sec_discussion}Discussion}

The phase diagram of the three phases, which are the step faceting (1D Bose solid + vacuum) phase, the step droplet (1D Bose fluid + 1D Bose solid) phase, and the GMPT (1D Bose fluid) phase (figure  \ref{phase_diagram}), is obtained by the PWFRG method \cite{pwfrg}, which is a variant of the DMRG method \cite{dmrg} (\S \ref{sec_phase_diagram} and \S \ref{sec_1Dattractive_bosons}).
The microscopic model is the p-RSOS model. 
The point-contact-type step-step attraction originates in the formation of a bonding state from overlapped orbitals at the meeting point of neighbouring steps. 

Our simple p-RSOS model provides a microscopic explanation of the type II anisotropy of the surface free energy described phenomenologically by Cabrera \cite{cabrera}.
This type II anisotropy results in two surfaces coexisting. 
Point-contact-type step-step attractive interactions are sufficient to cause this type II anisotropy.

For the vicinal surface inclined in the $\langle 101 \rangle$ direction, the step faceting phase does not appear because $r_f < 2r_{\rm ex}^{\langle 101 \rangle}$ (\S \ref{sec_exclusive_sphere}). 
The question arises as to whether the step faceting occurs if $r_f$ exceeds $2 r_{\rm ex}^{\langle 101 \rangle}$.
An example of this was given by the RSOS model coupled with the Ising model (RSOS-I model) \cite{akutsu03, akutsu01}.
The RSOS-I model was used to study a vicinal surface with adsorption where the coverage of adsorbates was less than 1.
The RSOS-I model and the modified RSOS-I model \cite{akutsu09-2} were mapped to a 304-vertex model \cite{akutsu01}.

In the RSOS-I model, the interaction among adsorbates, which is described by an Ising model, mediates the interaction among steps, and causes short-range attraction \cite{akutsu03}.
Although the steps at a site hardly meet for the vicinal surface inclined in the $\langle 101 \rangle$ direction (\S \ref{sec_1dboson101}),  the effective meeting of neighbouring steps mediated by clusters of adsorbates occurs.
The short-range attraction mediated by adsorbates covers the  distance between neighbouring steps. 

Moreover, the RSOS-I model contains $T_{f,1}$ and $T_{f,2}$, and $T_{f,1}$ is the end of the first-order transition.
The shape exponent of the normal direction, $\theta_n$, near $p=0$ changes to $\theta_n=4/3$   on only $T_{f,1}$ \cite{akutsu03} (\S \ref{shape_exponent2}).

Studying the terrace width distribution may provide another method for investigating spinless boson $n$-mers with quasi-impenetrability.
As Einstein {\it et al.} \cite{einstein99,giesen2000,einstein2001} reported, the terrace width distribution of the GMPT system obeys the Wigner distribution, $P_{\rm terrace}(s)$, where $s$ is the width of a terrace.
Because of the strict impenetrability between neighbouring steps, the value of $P_{\rm terrace}(s)$ near $s=0$ is made to be small and  $P_{\rm terrace}(s)|_{s \rightarrow +0}=0$.

However, in the system with attractive step-step interactions, the boson $n$-mers are formed as local condensates in the step droplet phase or the GMPT phase.
Hence, an increase in $P_{\rm terrace}(s)$ for small $s$ is expected.
Sathiyanarayanan {\it et al.} \cite{einstein09} studied the terrace width distribution, $P_{\rm terrace}(s)$, based on the special lattice model with steps touching by using the Monte Carlo method.
They found that  $P_{\rm terrace}(s)$ increased for small $s$, which is consistent with our expectations.
A more detailed study of the step width distribution with step-step attraction is required.

figure  \ref{phase_diagram} shows the phase diagram for a 19-vertex model.
den Nijs and Rommelse showed that a 19-vertex model can be mapped to a 1D quantum spin system \cite{dennij89}.
A similar result for the change in the shape exponent at the critical temperature in \S \ref{shape_exponent2} was obtained in a 1D XXZ quantum spin system for the high-field magnetization \cite{okunishi99} \footnote{In the 1D XXZ system, the Andreev field corresponds to the magnetic field, and the surface gradient corresponds to the magnetization.}.

\section{\label{sec_conclusions}Conclusions}

For a vicinal surface inclined in the direction of 
$\langle 101 \rangle$, the surface shows purely 1D FF behavior and does not cause the step faceting.
For a vicinal surface of the  p-RSOS model inclined in the $\langle 111 \rangle$ direction, we summarize our results as follows.

The phase diagram for the step faceting phase, step droplet phase, and GMPT phase is constructed in the area $-2<\epsilon_{\rm int}/\epsilon<0$ and $0<\kBT/\epsilon<2$ (\S \ref{sec_phase_diagram}).

In the step droplet phase, anisotropic surface tension has a disconnected shape around the (111) surface (\S \ref{sec_z}, \S \ref{sec_calculation}).
In the step faceting phase, the surface tension around the (001) surface becomes metastable except for the (001) surface; namely, the (001) surface is in direct contact with the (111) surface.

The sticky character of the steps roughens the (001) surface (\S \ref{roughening}).
The line of the roughening transition temperature divides the GMPT phase and the step droplet phase into the GMPT-I, the GMPT-II, the step droplet-I, and the step droplet-II. 
In the limit of $\rho^{QP} \rightarrow 0$, the chemical  potential of the 1D Bose fluid-I is zero, however the chemical potential of the 1D Bose fluid-II is finite (\S \ref{sec_1DBose_solid}).

The lower PSL between the step faceting phase and the step droplet phase-II (\S \ref{psl_Tf2}) is given by the condition such that  $\gamma_1= \lim_{n \rightarrow \infty} \gamma_n/n$ ($\mu_0=\mu_c$) (\S\ref{psl_Tf2}).
The lower PSL is well approximated by the curve obtained from the Ising model calculation by the IPW method.

To study the surface thermodynamics around the (111) surface, negative steps are introduced (\S\ref{sec_negative_step}) corresponding to the vacancies in the Bose solid.
On the upper PSL between the GMPT phase-I (-II) and the step droplet phase-I(-II), $f_{eff}(q)$ has the form of 
\beq
f_{eff}(q)=f(0)+\gamma_{q,1} q + C_{q,\rm eff} q^4 + {\mathcal O}(q^5). \nonumber
\eeq

The shape exponent along the $\langle 110 \rangle$ direction normal to the facet edge, $\theta_n$,  around the (001) facet ($p \rightarrow 0$) equals 2 near $T_{f,2}$ in the step droplet phase (\S \ref{sec_ecs}).
For the (111) facet, on only the upper PSL ($T_{f,1}$) the shape exponent of the normal direction to the (111) facet edge, $\theta_n$, equals 4/3 (\S \ref{shape_exponent2}).
In the GMPT phase,  $\theta_n=3/2$ around the (001) and  (111) facets.

In the RSOS model ($\epsilon_{\rm int}=0$), the expanded form of the surface free energy with respect to $\rho$ (equation  (\ref{eqfit})) has a $\rho^4$ term, which arises from the meeting of steps (collision of 1D Bose particles, \S\ref{sec_1dboson111}).

For the coarse-grained vicinal surface, the point-contact-type step-step interaction has an effective force range of $r_f=a/\sqrt{2}$, where $a$ (=1) is the lattice constant of the square lattice.
Because of the quasi-impenetrability, a boson has an effective exclusive sphere with the radius  $r_{\rm ex}^{\langle 111 \rangle}=1/(2 \sqrt{2})$.
The magnitude relationship between $r_f$ and $2r_{\rm ex}^{\langle 111 \rangle}$  affects the form of the Hamiltonian for the transfer matrix (\S\ref{sec_exclusive_sphere}).

\section*{Acknowledgments}

The author is grateful to Prof. Takanori Koshikawa and Prof. Tomohei Sasada for continuous encouragement.
This work was supported by a Japan Society for 
the Promotion of Science (JSPS) KAKENHI Grant
number 25400413.

\appendix

\section{\label{p-expansion}$p$-expanded form of effective surface free energy \cite{akutsuJPCM11}}

Let us consider a vicinal surface with impenetrable $n$-merged steps.
The surface free energy $f_n(\vec{p})$ of the vicinal surface must satisfy the GMPT form
\beqa
f_n(\rho_n)&=& f(0)+\gamma_n  \rho_n + B_n   \rho_n^3 + C_{n}  \rho_n^4 \nonumber \\
&&+ D_n   \rho_n^5 + {\cal{O} }(\rho_n^6), \label{fn}\\
B_n&=&\pi^2 (\kBT)^2/[6 \tilde{\gamma}_n] \label{univ_rel}
\eeqa
where $\gamma_n$ is the step tension of an $n$-merged step, $B_n$ is the step interaction coefficient of an $n$-merged step, and $C_n$ and $D_n$ are coefficients.
The equation  (\ref{univ_rel}) represents the universal relationship \cite{akutsu88} for an $n$-merged step, where $\tilde{\gamma}_n$  is the step stiffness of an $n$-merged step defined by $\tilde{\gamma}_n=\gamma_n+\partial^2 \gamma_n/\partial \phi^2$. 

First, we expand these coefficients with respect to $n$ around $n=1$, considering $\rho_n=|\vec{p}|/d_n$, $d_n=n d_1$, $\gamma_n \sim n \gamma_1$, and $B_n \sim B_1/n$.
After some calculations, we obtain 
\beqa
f_n(|\vec{p}|)&=& f(0)+\left[\gamma_1    -\gamma_1^{(1)}   +\frac{\gamma_1^{(2)}   }{2} \right. \nonumber \\
&&+ (\gamma_1^{(1)}   -\gamma_1^{(2)}   )n 
\left. 
+ \frac{\gamma_1^{(2)}   }{2}n^2 \right] \frac{|\vec{p}|}{d_1} \nonumber \\
+ \frac{B_1   }{n^4} \frac{|\vec{p}|^3}{d_1^3}&& \left[1+ (n-1)\frac{\tilde{\gamma}_1^{(1)}   }{\tilde{\gamma}_1   }
+\frac{1}{2}(n-1)^2 \frac{\tilde{\gamma}_1^{(2)}   }{\tilde{\gamma}_1   }     
\right]^{-1}\nonumber \\
&&+ \frac{C_n   }{n^4}\frac{ |\vec{p}|^4}{d_1^4} + {\cal{O}} (|\vec{p}|^5), \label{fnp2}
\eeqa
where  $\gamma_1^{(m)}  $ and $\tilde{\gamma}_1^{(m)}  $ are defined by
\beq
\gamma_1^{(m)}   =\left. \frac{\partial^m (\gamma_n   /n)}{\partial n^m}\right|_{n=1}, \quad \tilde{\gamma}_1^{(m)}   =\left. \frac{\partial^m (\tilde{\gamma}_n   /n)}{\partial n^m}\right|_{n=1}. \label{eq_gamma_expand}
\eeq

Recalling that the state with boson $n$-mers is not an eigenstate, we consider the mixture of boson $n$-mers with various values of $n$.
Then, we replace $n$ with $\langle n \rangle$.

Next, we expand $\langle n \rangle$ with respect to $|\vec{p}|$ around $|\vec{p}|=0$.
Namely,
\beqa
\langle n  \rangle &=& 1+ n_0^{(1)}   |\vec{p}|+ \frac{1}{2}n_0^{(2)}   |\vec{p}|^2+ \frac{1}{6}n_0^{(3)}   |\vec{p}|^3 + \cdots \nonumber \\
 n_0^{(m)}  &=& \left.\frac{\partial^m \langle n   \rangle }{\partial |\vec{p}|^m}\right|_{|\vec{p}|=0+} . \label{n_expand}
\eeqa

Finally, substituting equation  (\ref{n_expand}) into equation  (\ref{fnp2}), and after some manipulation, the effective vicinal surface free energy can be expressed as 
\beqa
f_{\rm eff}(\vec{p}) \equiv f_{\langle n \rangle}(\vec{p})=f(0)+\gamma_1    \frac{|\vec{p}|}{d_1}+A_{\rm eff}    |\vec{p}|^2 \nonumber \\
+B_{\rm eff}    |\vec{p}|^3 +C_{\rm eff}    |\vec{p}|^4 +{\cal O} (p^5), \label{fpeff}
\eeqa
where
\beqa
A_{\rm eff}   &=&n_0^{(1)}   \gamma_1^{(1)}   /d_1 \label{eqAeff} \\
B_{\rm eff}   &=&  \frac{1}{2d_1} \left[ n_0^{(2)}   \gamma_1^{(1)}   + (n_0^{(1)} ) ^2 \gamma_1^{(2)}   \right] + \frac{B_1   }{d_1^3}   \label{eqBeff} \\
C_{\rm eff}   &=& \frac{1}{6d_1}\left[ n_0^{(3)}   \gamma_1^{(1)}    +  3 n_0^{(1)}   n_0^{(2)}   \gamma_1^{(2)}     \right] \nonumber \\
&&- \frac{B_1    n_0^{(1)}   }{d_1^3}  \left( 4  +  \frac{\tilde{\gamma}_1^{(1)}   }{\tilde{\gamma}_1   }\   \right) 
+ \frac{C_1   }{d_1^4}  . \label{eqCeff}
\eeqa

\section{\label{sec_correspond}Ground state energy}

\subsection{\label{sec_ff} Fermion-like behavior }


\begin{figure}
\centering
\includegraphics[width=5 cm,clip]{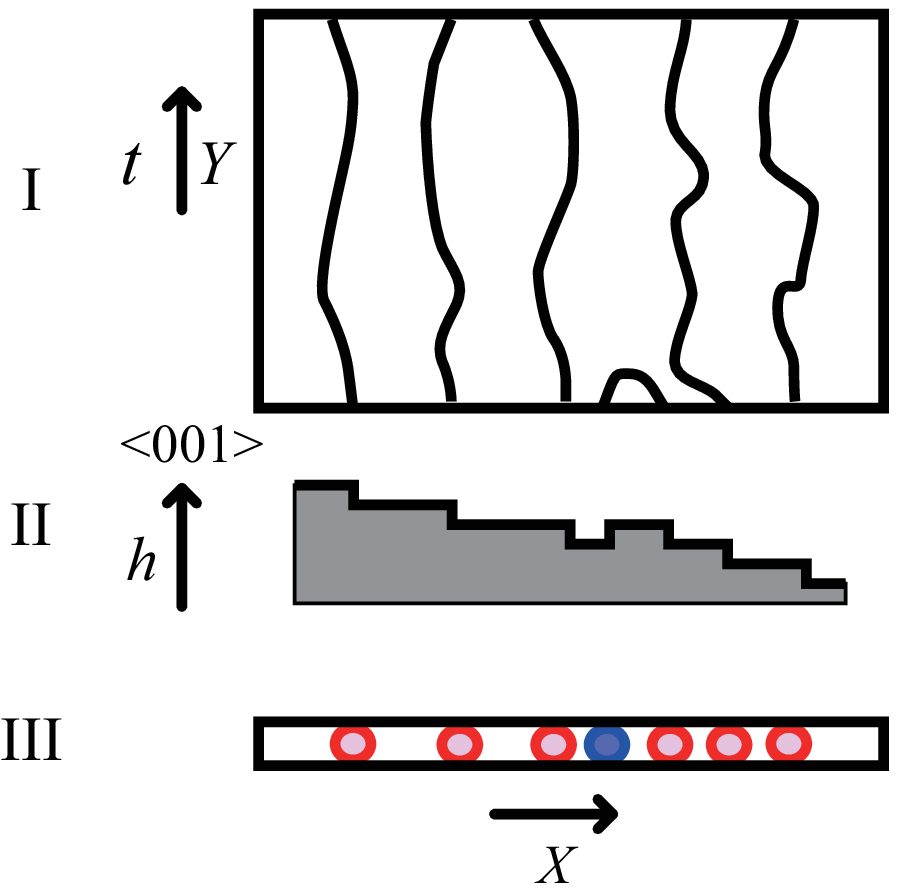}%
\caption{
Relationship between a vicinal surface and 1D quantum particles.
I: Top view of a coarse-grained vicinal surface \cite{kogut,suzuki}.
II: Side view of the vicinal surface of I.
III: Image of 1D quantum particles corresponding to II.
Part of the surface island in I is described by a particle-antiparticle pair.  The antiparticle is shown in blue.
}
\label{vicinalTSK}
\end{figure}

We used the transfer matrix method to calculate the partition function.
The transfer matrix can be translated into the Hamiltonian of the quantum particles in a 1D lattice with the path integral method\cite{kogut}.
By treating the mean step-running direction as the imaginary time direction (figure  \ref{vicinalTSK} I), the two-dimensional (2D) surface system can be mapped to a 1D quantum particle system. 
In the mapping, a step is converted to a trajectory or a path of a particle, and an island is converted to the pair-creation and pair-annihilation of the particle and an antiparticle (figure  \ref{vicinalTSK} I--III).
Alternatively, the RSOS model can be mapped to a system of spinless quantum particles on a 1D lattice at zero temperature $T^{\rm QP}=0$ by applying the Suzuki-Trotter formula \cite{suzuki} to the product of the transfer matrices in the mean running direction of the surface steps.

Let us consider a coarse-grained RSOS system with steps of size $L$ (space direction) $\times M$ (time direction).
Treating the steps as linear elementary excitations with strict impenetrability in two dimensions, the transfer matrix for the continuous system is written as $\exp [- \beta {\cal H}^{\rm (FF)} (\sigma)$, where $\beta=1/\kBT$, and ${\cal H}^{\rm (FF)} (\sigma)$ is written as
\beqa
{\cal H}^{\rm (FF)} (\sigma)&=&Lf(0)+ \int_0^L {\rm d}x \  [\mu_0 \phi^{\dagger} (x) \phi(x) \nonumber \\
&&+\frac{\sigma}{2} \ \partial \phi^{\dagger} (x) \partial \phi(x)].
\label{hamilFF}
\eeqa
where $\phi^{\dagger}(x)$ and $\phi(x)$ are the fermion operators and $\mu_0$ is the chemical potential of a fermion with a low-density limit.
After some calculations \cite{akutsu88}, the ground state energy density for 1D FF $E^{FF}$ is obtained as
\beqa
E^{\rm FF}(\rho^{\rm QP})&=& f(0)+ \mu_0 \rho^{\rm QP} 
+ \frac{\sigma \pi^2}{6} (\rho^{\rm QP})^3 \nonumber \\ 
&&
 + {\cal O}((\rho^{\rm QP})^5),\label{ground_eFF}
\eeqa
where $\rho^{\rm QP}=n^{\rm QP}/L$ is the density of quantum particles and $n^{\rm QP}$ is the number of quantum particles.
Equation (\ref{ground_eFF}) consists of odd powers of $\rho^{\rm QP}$.
Comparing the ground state energy, equation  (\ref{ground_eFF}), with the surface free energy, equation  (\ref{f-gmpt}), we have $\mu_0=\gamma$ and $\sigma \pi^2/6=B$.

\subsection{\label{sec_boson}Boson-like behavior}

Let us consider the Hamiltonian of a 1D spinless Bose gas in a continuous system as 
\beqa
{\cal H}^{\rm (Bose)} (\sigma,c)=\int {\rm d}x \  [\mu_0 \phi^{\dagger} (x) \phi(x) \nonumber \\
+\sigma \ \partial \phi^{\dagger} (x) \partial \phi(x) 
+c \ \phi^{\dagger}(x) \phi^{\dagger} (x) \phi(x)\phi(x)],
\label{hamilLieb}
\eeqa
where $\phi^{\dagger}(x)$ and $\phi(x)$ are the Boson operators, $\mu_0$ is the chemical potential of a boson at the low-density limit, and $c$ is the two-body collision coefficient.
The ground state energy density for $c>0$ (repulsive) is given exactly by Lieb-Liniger \cite{lieb} by solving the Bethe ansatz integral equation.
By using the boson density, $\rho^{\rm QP}$, the ground state energy density of the Bose gas system is written  as \cite{okunishi99}
\beqa
E^{\rm Bose}(\rho^{\rm QP})&=& \mu_0 \rho^{\rm QP} 
+ \frac{\sigma \pi^2}{3} (\rho^{\rm QP})^3 
- \frac{4 \sigma^2 \pi^2}{3c} (\rho^{\rm QP})^4 \nonumber \\ 
&&+ {\cal O}((\rho^{\rm QP})^5).\label{ground_eBose}
\eeqa

Comparing equation  (\ref{ground_eBose}) with equation  (\ref{ground_eFF}) shows that the term $(\rho^{\rm QP})^4$ gives the difference between the FF system and the Bose system.

Equation (\ref{ground_eBose}) provides the approximate form for the surface free energy.
Comparing equations (\ref{eqfit}) and (\ref{ground_eBose}) gives $\gamma=\mu_0$, $B=\sigma\pi^2/3$, and $C_4= -4\sigma^2\pi^2/(2c)$.
Using these relationships, we obtain $\sigma$ and $c$ from $B$ and $C_4$. \\

{\bf References}\\


\end{document}